\documentclass[twocolumn]{article}
\usepackage{times}
\usepackage{graphics}
\newcommand{\bc}{\begin{center}}
\newcommand{\ec}{\end{center}}
\newcommand{\be}{\begin{equation}}
\newcommand{\ee}{\end{equation}}
\newcommand{\ba}{\begin{array}}
\newcommand{\ea}{\end{array}}
\newcommand{\bea}{\begin{eqnarray}}
\newcommand{\eea}{\end{eqnarray}}
\newcommand{\noi}{\noindent}
\def\ga{\mathrel{\mathchoice {\vcenter{\offinterlineskip\halign{\hfil
$\displaystyle##$\hfil\cr>\cr\sim\cr}}}
{\vcenter{\offinterlineskip\halign{\hfil$\textstyle##$\hfil\cr
>\cr\sim\cr}}}
{\vcenter{\offinterlineskip\halign{\hfil$\scriptstyle##$\hfil\cr
>\cr\sim\cr}}}
{\vcenter{\offinterlineskip\halign{\hfil$\scriptscriptstyle##$\hfil\cr
>\cr\sim\cr}}}}}

\def\bitem#1\par{\noindent\hangindent1.0\parindent\hangafter=1\rm#1\par\smallskip}
\begin{document}
\twocolumn[
\phantom{.}
\vspace{-0.5cm}
\bc
{\large\bf
SCATTERING AND POLARIZATION PROPERTIES \\ OF THE \\
           NON-SPHERICAL PARTICLES}

\baselineskip=11.5pt
\vspace{0.5cm}
{Nikolai V. Voshchinnikov}\\
Astronomy Department and Sobolev Astronomical Institute, St. Petersburg University, \\
198504 St. Petersburg-Peterhof, Russia
\ec
]
\phantom{.}
\vspace{0.1cm}
\baselineskip=11.5pt

\bc{\bf ABSTRACT}\ec

 The albedo and dichroic polarization
of spheroidal particles are studied.
For the description of the phase  function
two asymmetry parameters $g_{||}$ and $g_{\bot}$
characterizing the anisotropy in forward/backward and
left/right directions are introduced and calculated.
The consideration is based on the solution to the light scattering
problem by the Separation of Variables Method
(Voshchinnikov and Farafonov, 1993).

\bigskip
\bc{\bf 1.\  INTRODUCTION}\ec

In many scientific and engineering applications
prolate and oblate spheroids are appropriate  models for real particles.
We consider the light scattering by homogeneous
spheroids using the Separation of Variables Method (SVM).
 The optical properties of prolate  spheroids of various aspect ratios $a/b$
 for several refractive indices $m$ are calculated
and the results for the particles of the same volume are compared.

\bigskip
\bc{\bf 2.\  GENERAL DEFINITIONS AND METHOD}\ec

A spheroid (ellipsoid of revolution) is obtained by the rotation of
an ellipse around
its major axis (prolate spheroid) or its minor axis (oblate spheroid).
 The ratio of the major semiaxis $a$ to the minor semiaxis $b$
 (i.e. the aspect ratio $a/b$)
characterizes the particle shape which may vary from a nearly spherical one
($a/b \approx 1$) to a needle or a disk ($a/b \gg 1$).
\\

We assume that an incident plane wave
has the  wavelength $\lambda$. Let $\alpha$ denote the angle
between the propagation direction and the rotation axis of the spheroid 
($0^{\circ} \leq \alpha \leq 90^{\circ}$).
\\
 
For the axial propagation ($\alpha=0^{\circ}$),  there is
no polarization of transmitted radiation due to symmetry.
 If $\alpha \ne 0^{\circ}$, two  cases
of polarization of the incident radiation have to be considered:
the electric vector $\vec{E}$
is parallel (TM mode) or perpendicular (TE mode) to the plane defined
by the spheroid's rotation axis and the wave propagation vector.
\\

\newpage
\phantom{.}
\vspace{0.4cm}

The size parameter is given by
\be
x_{\rm V} = \frac{2\pi r_{\rm V}}{\lambda},
\ee
where $r_{\rm V}$ is the radius of the sphere whose volume is
equal to that of the spheroid.
 The radius $r_{\rm V}$ for prolate  spheroids  is defined as
\begin{equation}
r_{\rm V}^{3} = a b^{2}\,.
\end{equation}

One usually calculates the efficiency factors $Q = C/G$
which are the ratio of the corresponding cross-sections $C$ to
the geometrical cross-section $G$ of the prolate spheroid
(the area of the particle's shadow)
\begin{equation}
G(\alpha) = \pi b \left(a^2\sin^2\alpha
            + b^2\cos^2\alpha\right)^{1/2}\,.
\end{equation}

In order to compare the optical properties of the particles of different
shapes it is convenient to consider the ratios of the cross-sections
for spheroids to the geometrical cross-sections of the
equal volume spheres, $C/\pi r^{2}_{\rm V}$.
For a prolate spheroid they can be found as
\be
\frac{C}{\pi r^{2}_{\rm V}} = \frac{[(a/b)^2\sin^2\alpha
                             +\cos^2\alpha]^{1/2}}{(a/b)^{2/3}} Q \,.
\ee

The albedo of a particle can be calculated from the extinction and
scattering cross-sections
\be
\Lambda = \frac{C_{sca}}{C_{ext}}\,.
\label{lam}
\ee

\begin{figure*}[htb]
\resizebox{175mm}{!}{\includegraphics{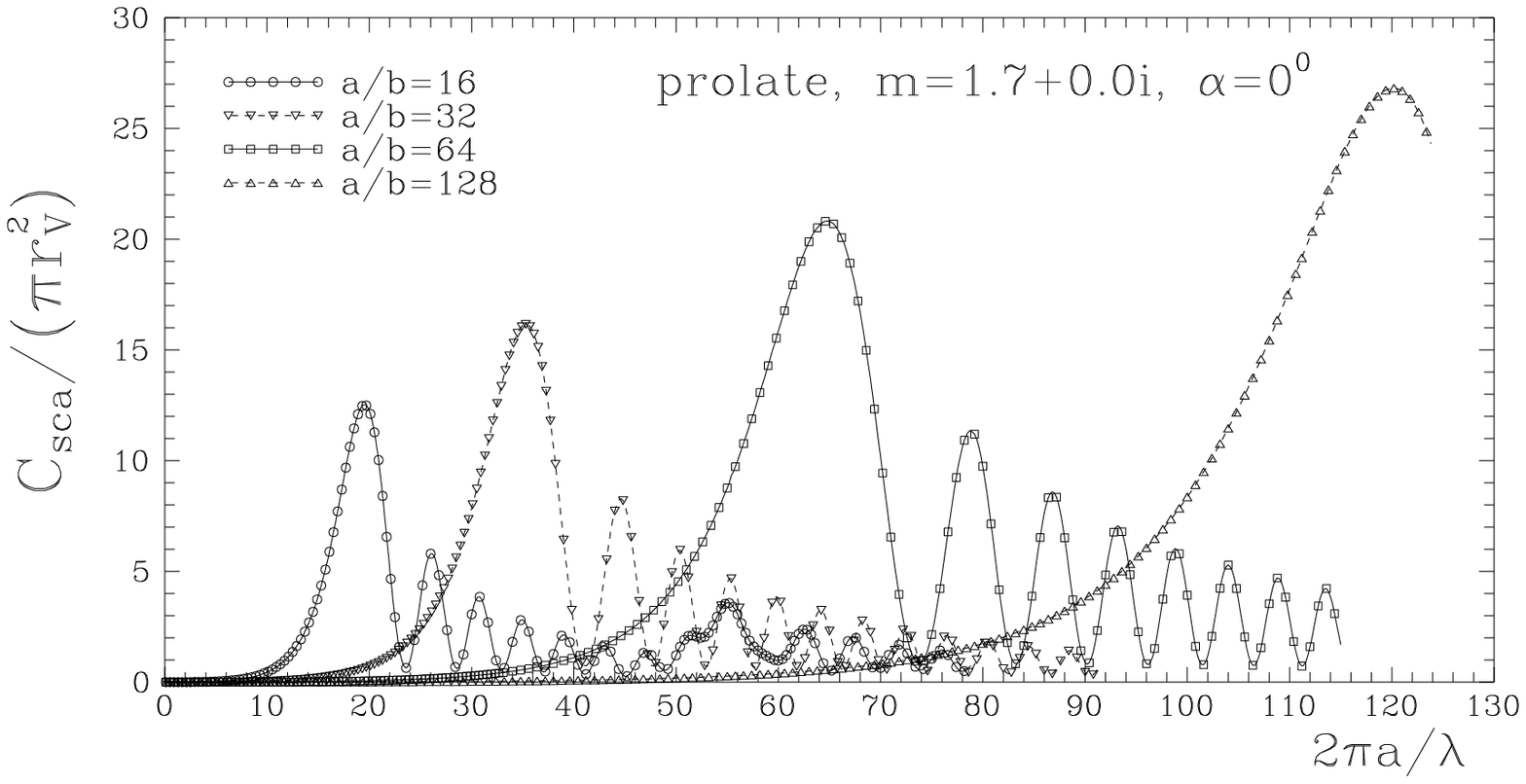}}
\caption[]{
Normalized scattering cross-sections for elongated
spheroidal particles in dependence on size parameter $2\pi a / \lambda$.
Transformation to the parameter $x_{\rm V}$ is given by
the expression $x_{\rm V}= (a/b)^{-2/3} (2\pi a / \lambda)$}
\label{f1}
\end{figure*}

In general, the radiation scattered by aligned spheroidal particles
has an azimuthal asymmetry that provokes a non-coincidence of
the directions of the radiation pressure force
and of the wave-vector of incident radiation
(Voshchinnikov, 1990; Il'in and Voshchinnikov, 1998).
Another consequence of the azimuthal asymmetry is the anisotropy of
the phase function in the  left/right direction.
The geometry of the phase function
in forward/backward and left/right directions
may be characterized by two asymmetry parameters  $g_{||}$
and $g_{\bot}$, respectively. Expressions for them can be found
from the consideration of radiation pressure
(Voshchinnikov, 1990; Il'in and Voshchinnikov, 1998)
\be
g_{||} =
\frac{{\cal A} \left(m, x_{\rm V}, a/b, \alpha \right) \cos \alpha
+ {\cal B} \left(m, x_{\rm V}, a/b, \alpha \right) \sin \alpha}
{Q_{\rm sca}\left(m, x_{\rm V}, a/b, \alpha \right)}\,,
\ee
\be
g_{\bot}=
\frac{{\cal A} \left(m, x_{\rm V}, a/b, \alpha \right) \sin \alpha
- {\cal B} \left(m, x_{\rm V}, a/b, \alpha \right) \cos \alpha}
{Q_{\rm sca}\left(m, x_{\rm V}, a/b, \alpha \right)}\,,
\ee
where $Q_{\rm sca}$ is the scattering efficiency factor,
and the coefficients ${\cal A}$ and ${\cal B}$
are
\be
{\cal A} = {\cal K} \left(x_{\rm V}, a/b, \alpha \right)
\int_{0}^{2\pi}\int_{0}^{\pi}i(\theta,\varphi)\cos \theta \sin \theta
        {\rm d}\theta {\rm d}\varphi,
\label{A}
\ee
\be
{\cal B} = {\cal K} \left(x_{\rm V}, a/b, \alpha \right)
\int_{0}^{2\pi}\int_{0}^{\pi}i(\theta,\varphi)\sin^2 \theta \cos \varphi
        {\rm d}\theta {\rm d}\varphi.
\label{B}
\ee
In Eqs.~(\ref{A})~--~(\ref{B}), ${\cal K}$ is a parameter,
$i(\theta,\varphi)$ the dimensionless intensity of scattered
radiation (phase function).
From a symmetry consideration, it is clear that for spheroids
$g_{\bot}= 0$ if $\alpha = 0^{\circ}$ or $90^{\circ}$.
Note that in all cases the following inequality is valid
\be
-1 \leq  \sqrt{g_{||}^2 + g_{\bot}^2} \leq 1 \,.
\ee

The dichroic polarization efficiency is defined
by the extinction cross-sections for TM and TE modes
\be
\frac{P}{\tau} =  \frac{C^{\rm TM}_{ext}-C^{\rm TE}_{ext}}
                          {C^{\rm TM}_{ext}+C^{\rm TE}_{ext}} \cdot 100\% \,.
\label{pol}
\ee
This ratio describes the efficiency to  polarize light
transmitted through an uniform slab consisting of non-rotating
particles of the same orientation.
\\


The optical properties of spheroidal particles can be determined by various
methods of light scattering theory
(see Mishchenko et al., 2000 for a review).
We use the SVM's solution developed by Farafonov and numerical
code based on it
(see Voshchinnikov and Farafonov, 1993 for more details).
A comparison of different
numerical codes and  benchmark results can be found in the
paper of Voshchinnikov et al.~(2000).
\\

The main problem of the SVM for spheroids is the difficulties with
computations of the spheroidal wavefunctions. Especially it is related to
very elongated particles with sizes larger than the wavelength because
the standard expansions of the prolate spheroidal wavefunctions
in series of the Legendre functions do not converge.
Farafonov and Voshchinnikov (2000, in preparation)
have considered new expansion  of the prolate wavefunctions that
opens a possibility to calculate the optical properties
of particles with $a/b \gg 1$.
Firstly, this method was applied to calculations of the radial wavefunctions
with the index $m=1$ that allows us to study the case of axial propagation
of radiation. Some results are shown in Fig.~\ref{f1} for non-absorbing
spheroids with the refractive index $m=1.7+0.0i$.
 As the extinction and scattering
cross-sections were the same (with 6 and more digits),
the values presented are expected to be correct.

\bigskip
\bc{\bf 3.\ NUMERICAL RESULTS}\ec

We present some results illustrating
the behaviour of the optical properties of prolate spheroidal particles
in a fixed orientation.
\\

\newpage
\noi{\bf 3.1 Albedo}
\\

\begin{figure}[h]
\resizebox{82mm}{!}{\includegraphics{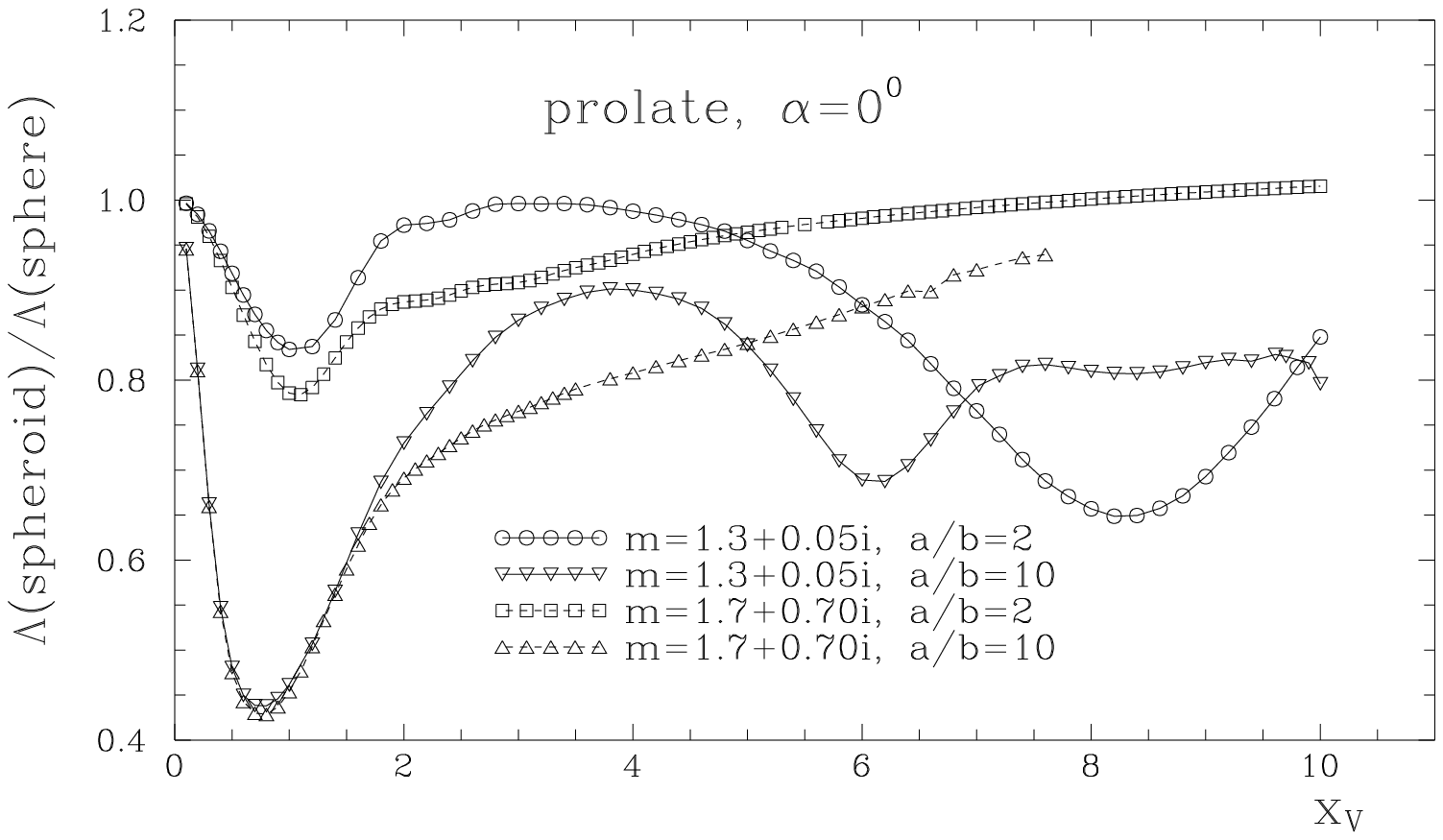}}
\caption[]{
Albedo of spheroidal particles
normalized relative to albedo of spherical particles with
the same refractive index
}
\label{f2}
\end{figure}

The integral scattering properties of particles are characterized
by their albedo. This quantity depends
on the particle size and, in general, on the particle shape.
 Figure~\ref{f2} shows the size dependence of the albedo for spheroids with
$m=1.3+0.05i$ and $m=1.7+0.7i$ and the aspect ratios $a/b = 2$ and 10.
The calculations were made for prolate particles
and $\alpha = 0^{\circ}$ (we adopt that the incident radiation is non-polarized).
In this  case in comparison with others
(oblate spheroids, oblique incidence of radiation) the largest deviations
of the ratio $\Lambda (\rm spheroid)/\Lambda (\rm sphere)$ from unity occur
(see Voshchinnikov et al., 2000 for discussion).

It is seen that {\it the albedo for large non-spherical particles becomes close to
that of spheres.}
Our calculations made for  particles with different absorption
show that the distinction of the albedo for spheres and spheroidal particles
remains rather small (within $\sim 20\,\%$) if the ratio
of the imaginary part of the refractive index to its real part
$k/n \ga 0.2 - 0.3$.
\\

\noi{\bf 3.2 Asymmetry parameter}
\\

\begin{figure}[h]
\resizebox{62mm}{!}{\includegraphics{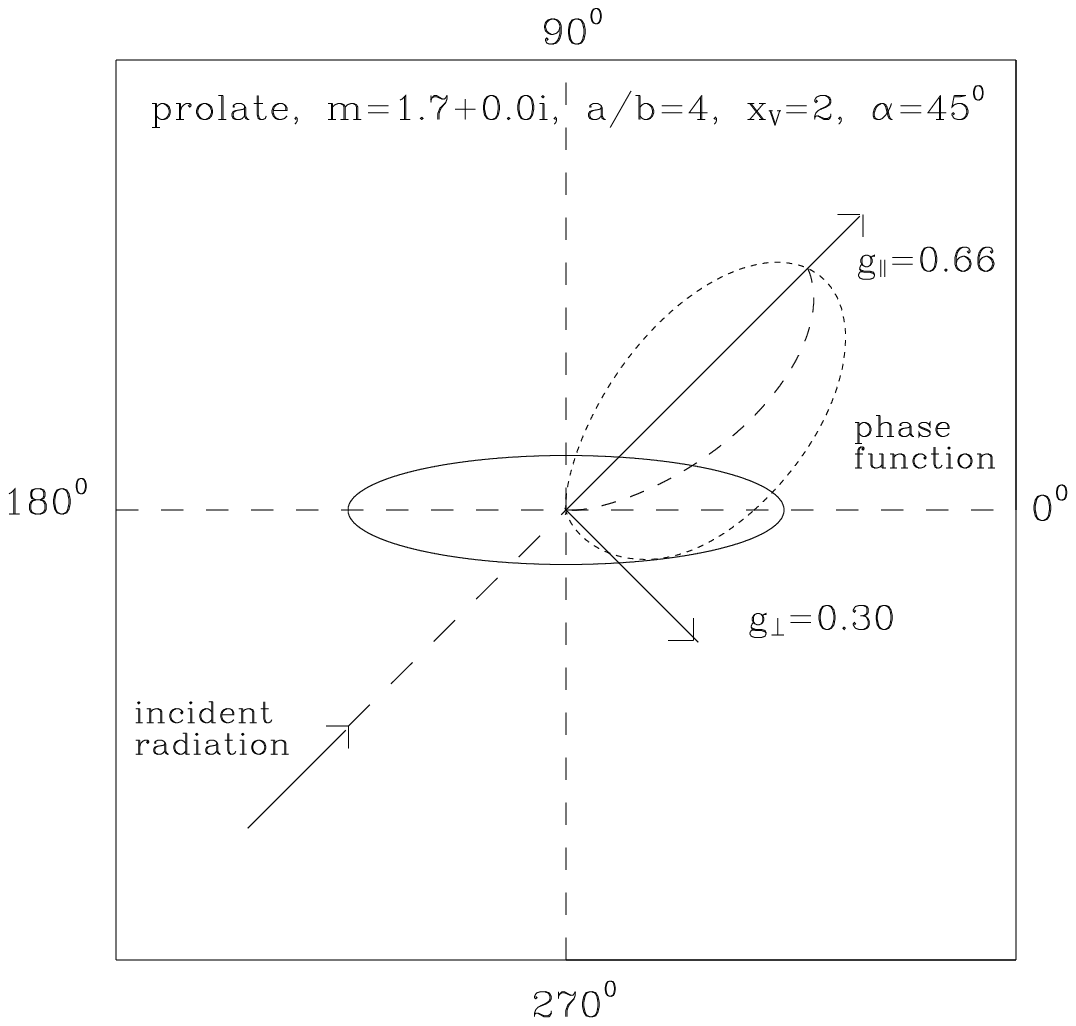}}
\caption[]{Geometry of light scattering by a prolate spheroid.
The wave-vector of incident non-polarized radiation
forms the angle $\alpha = 45^{\circ}$ with the rotation axis of
the spheroid.
Short-dashed curve shows the angular distribution of
the scattered radiation (phase function).
The part of scattered radiation which is symmetric relative to the direction
of incident radiation is plotted by long-dashed curve.
The  radial and transversal asymmetry factors
and their values are indicated}
\label{fff}
\end{figure}

Another characteristic of scattered radiation is the
asymmetry parameter describing the spatial distribution
of scattered radiation around a particle.
Usually the anisotropy in forward/backward direction
is only considered. However, the radiation scattered by
any aligned non-spherical particle
possesses also an anisotropy in the right/left direction
in the case of oblique incidence of radiation (see Fig.~\ref{fff}).
\bc
\ec

\begin{figure}[htb]
\resizebox{72mm}{!}{\includegraphics{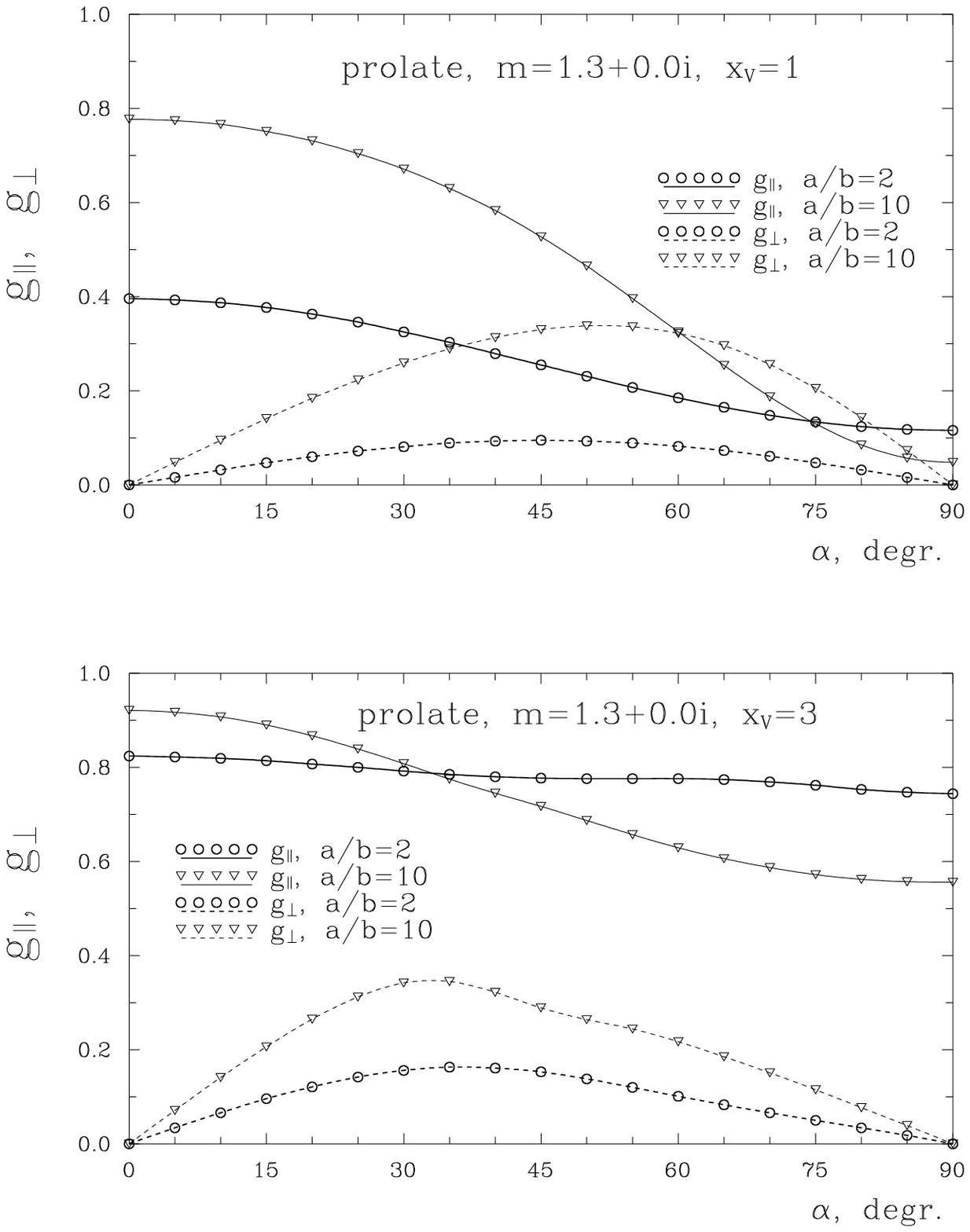}}
\caption[]{
Angular dependence of the asymmetry factors of scattered radiation
}
\label{f4}
\end{figure}

As it is seen from Fig.~\ref{f4} both asymmetry factors
$g_{||}$ and $g_{\bot}$ change with $a/b$ and $\alpha$.
The values of radial asymmetry factor $g_{||}$ decrease
with a growth of $\alpha$ when the path of radiation reduces
from $2a \ (\alpha=0^{\circ})$ to $2b \ (\alpha=90^{\circ})$.
The transversal asymmetry factor $g_{\bot}$ can be rather
large and even exceeds the radial one.
Because the geometry of light scattering by very elongated spheroids
approaches  that of infinite cylinders,\footnote{In this
case the scattered radiation forms the conical surface
with the opening angle $2 \alpha$.}
{\it such particles scatter more radiation ``to the side"
than in forward direction.}
\begin{figure}[htb]
\resizebox{72mm}{!}{\includegraphics{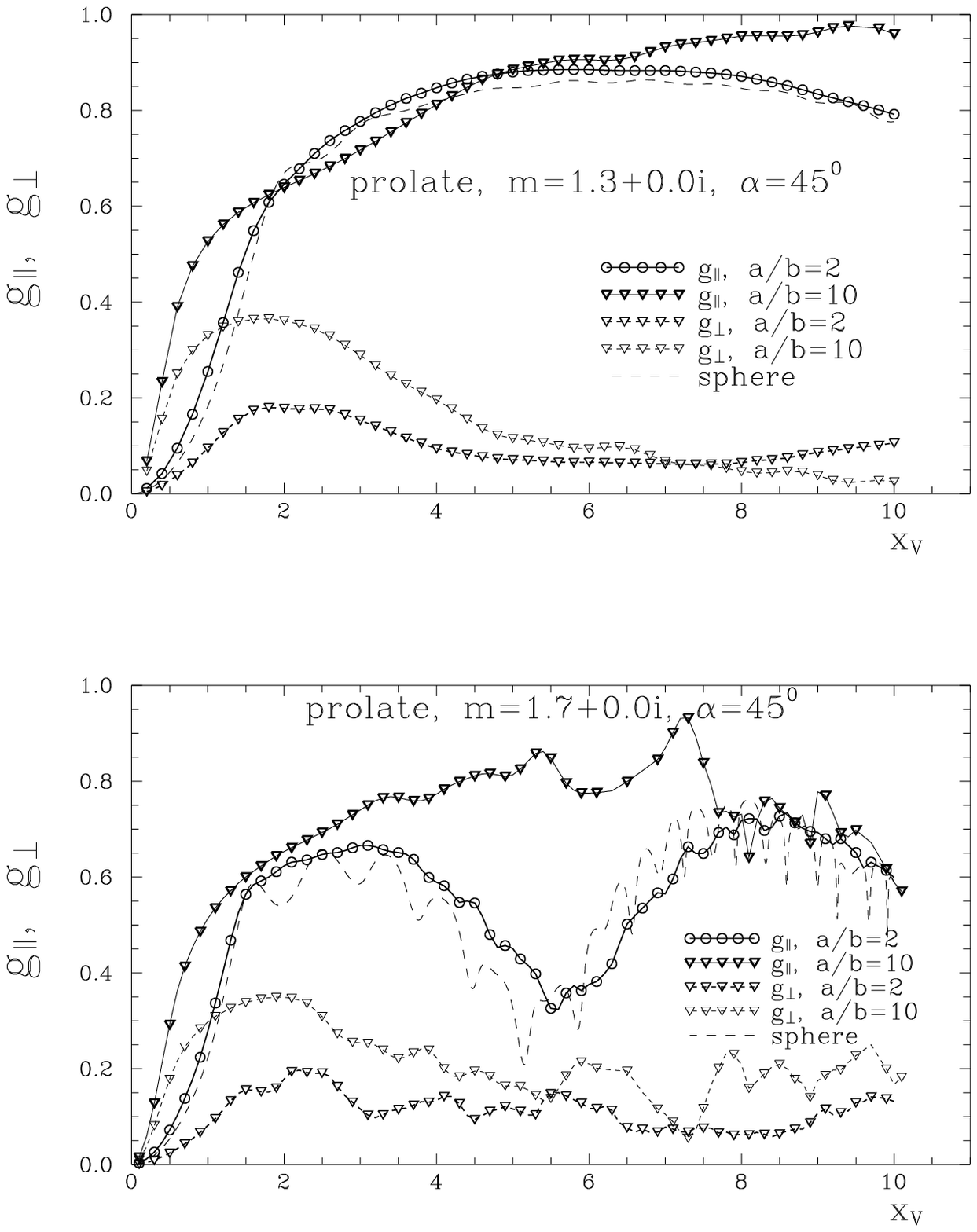}}
\caption[]{
Size dependence of the asymmetry factors of scattered radiation
}
\label{f5}
\end{figure}

The size dependence of asymmetry factors is plotted in Fig.~\ref{f5}.
It shows that the variations of $g_{||}(x_{\rm V})$ for spheroids
with $a/b=2$ are rather similar to those of spheres.
However, the radiation scattered by spheroids has a
noticeable azimuthal dependence which is absent for spherical particles.
If $x_{\rm V} \ga 4$ the azimuthal anisotropy of scattered radiation
reduces and  $g_{\bot}$ drops.
\\

\noi{\bf 3.3 Polarization}
\\

If a volume contains aligned non-spherical particles,
the initially non-polarized incident radiation
will be partially polarized after having passed the volume.
The simplest and at the same time extreme
case of particles' alignment is the perfect alignment of non-rotating
particles (picked fence orientation).
 The maximum polarization usually occurs when
the major axes of the particles are
perpendicular to the direction of the incident radiation
($\alpha = 90^{\circ}$).
\\

\begin{figure}[htb]
\resizebox{82mm}{!}{\includegraphics{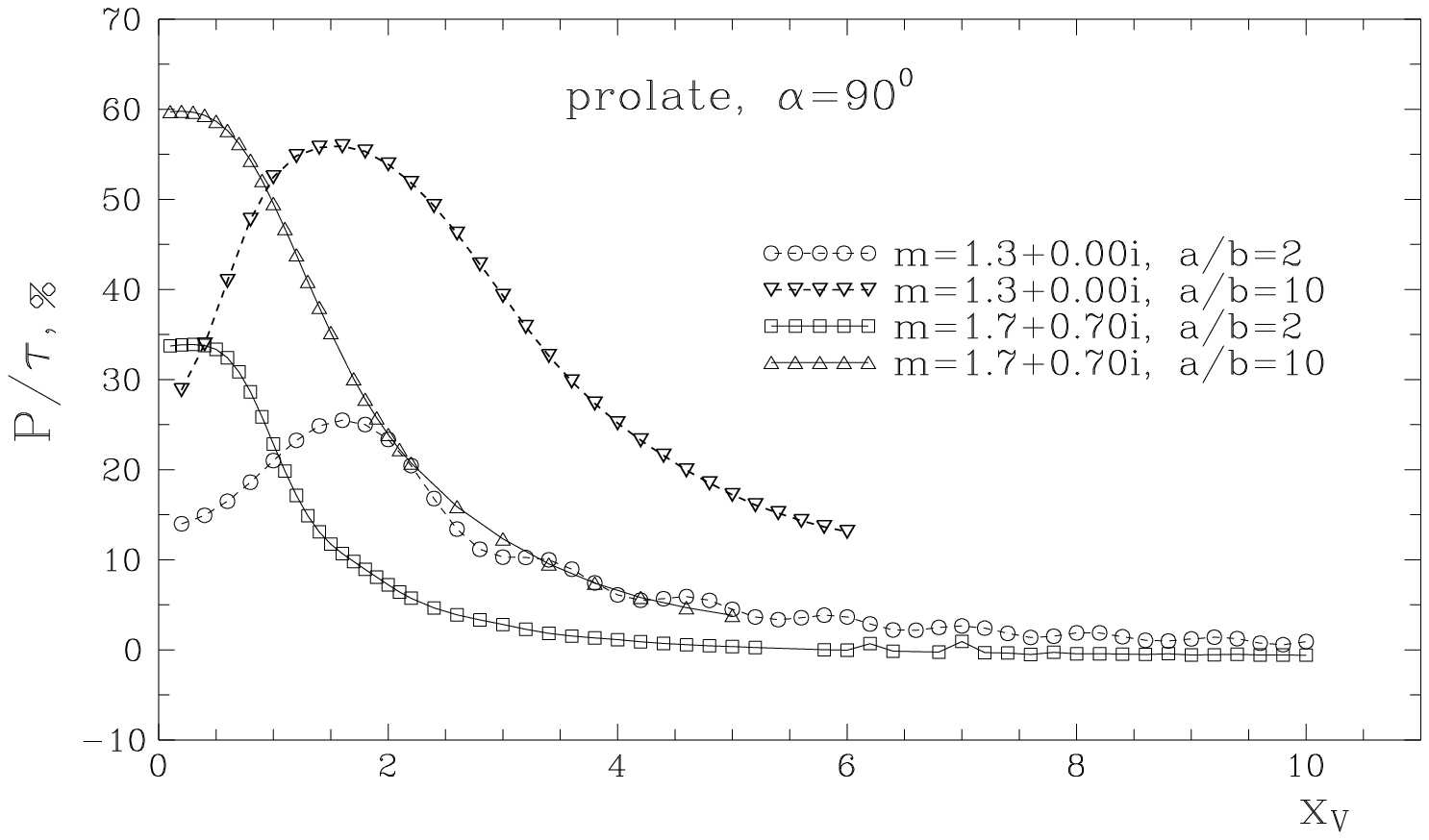}}
\caption[]{
Size dependence of the polarization efficiency
}
\label{f3}
\end{figure}

The behaviour of the polarization efficiency $P/\tau$ 
for non-absorbing and absorbing spheroids is shown in Fig.~\ref{f3}.
 It is clearly seen that a relatively large particles produce
{\it no polarization independent of their shape}.
 For absorbing particles, it occurs at smaller $x_{\rm V}$ values
than for non-absorbing particles.
 This effect should  depends on the imaginary part
of the refractive index as well.

\bigskip
\bc{\bf 4.\ CONCLUSIONS}\ec

The albedo of large non-spherical particles
exhibits only a weak dependence on the particle shape
if the ratio of the imaginary part of the refractive index to its real part
$k/n \ga 0.2 - 0.3$.
\\

The radiation scattered by aligned spheroidal particles
has an azimuthal asymmetry and its geometry may be described by
two asymmetry parameters  $g_{||}$ and $g_{\bot}$ showing the deviations
from the symmetric scattering
in forward/backward and left/right directions, respectively.
The transversal asymmetry factor $g_{\bot}$ can be rather
large and even exceeds the radial one, therefore,
very elongated spheroids scatter more radiation ``to the side"
than in forward direction.
\\

Particles larger than a certain minimum size
do not polarize the transmitted radiation independent of their shape.

\bigskip
\bc {\bf ACKNOWLEDGEMENTS} \ec

The author is thankful to V.G.~Farafonov for discussion and
to V.B.~Il'in for valuable comments.
This work was financially supported by the INTAS foundation
(grant Open Call 99/652).
\\

\newpage
\bc{\bf REFERENCES}\ec

\bitem{Il'in, V.B., and  N.V. Voshchinnikov, 1998:
Radiation pressure on non-spherical dust grains
in envelopes of late-type giants.
{\it Astron. Astrophys. Suppl.}, {\bf 128}, 187-196.}

\bitem{Mishchenko, M.I., J.W. Hovenier, and L.D. Travis (eds.), 2000:
         {\it Light Scattering by Nonspherical Particles}.
         Academic Press, 690 pp.}

\bitem{Voshchinnikov, N.V., 1990:
         Radiation pressure on spheroidal particles.
          {\it Soviet Astronomy}, {\bf 34}, 429-432.}


\bitem{Voshchinnikov, N.V., and V.G. Farafonov, 1993:
Optical properties of spheroidal particles.
  {\it Astrophysics and Space Science}, {\bf 204}, 19-86.}

\bitem{Voshchinnikov, N.V., V.B. Il'in, Th. Henning, B. Michel,
      and V.G. Farafonov, 2000:
      Extinction and polarization of radiation by absorbing
      spheroids: shape/size effects and some benchmarks.
      {\it Journal of Quantitative Spectroscopy and Radiative Transfer},
       {\bf 65}, 877-893.}

\end{document}